# Optimality of Moore neighborhoods in protein contact maps

Susan Khor
slc.khor@gmail.com

**Abstract**

A protein contact map is a binary symmetric adjacency matrix capturing the distance relationship between atoms of a protein. Each cell $(i, j)$ of a protein contact map states whether the atoms (nodes) $i$ and $j$ are within some Euclidean distance from each other. We examined the radius one Moore neighborhood surrounding each cell $(i, j)$ where $j > (i + 2)$ in complete protein contact maps by mutating them one at a time. We found that the particular configuration of a neighborhood is generally (97%) optimal in the sense that no other configuration could maintain or improve upon existing local and global efficiencies of the nodes residing in a neighborhood. Local efficiency of a node is directly related to its clustering measure. Global efficiency of a node is inversely related to its distance to other nodes in the network. This feature of the Moore neighborhood in complete protein contact maps may explain how protein residue networks are able to form long-range links to reduce average path length while maintaining a high level of clustering throughout the process of small-world formation, and could suggest new approaches to protein contact map prediction. Effectively, the problem of protein contact map prediction is transformed to one of maximizing the number of optimal neighborhoods. By comparison, Moore neighborhoods in protein contact maps with randomized long-range links are less optimal.

## 1. Motivation

A protein contact map is a binary symmetric adjacency matrix capturing the distance relationship between atoms of a protein. Each cell $(i, j)$ of a protein contact maps state whether the atoms (nodes) $i$ and $j$ are within some Euclidean distance from each other. Much has been written about the small-world nature of protein contact maps (PCM) or the protein residue networks (PRN) they represent [e.g.: 1-3], but less is known about how these particular small-worlds form. A random approach such as the canonical Watts' rewiring algorithm [4] would not work, not only because of physical and chemical restrictions, but in other purely descriptive ways too. Nonetheless, we made an attempt in [5] for comparison.

We observed the tendency for PRNs to maintain their initial level of clustering throughout the small world formation process. That is, as long-range links are added in monotonically increasing sequence distance order to a PRN with only short-range links initially, the mean path length decreases, but the mean clustering does not decrease much. This is in agreement with small-world formation for small $p$ [4]. But the high level of clustering reduces the speed at which mean path length decreases in PCMs. Fig. 1 demonstrates this network behavior for the much studied 2CI2 protein. Links are added in increasing sequence distance order. Given two edges *e1*$(p, q)$ and *e2*$(r, s)$, *e1* is shorter than *e2* in terms of sequence distance iff $|p - q| < |r - s|$. Fig. 2 compares the local and global efficiencies of initial PRNs (with only SE or short-range links), and the complete PRNs (with all links added) for 29 proteins.

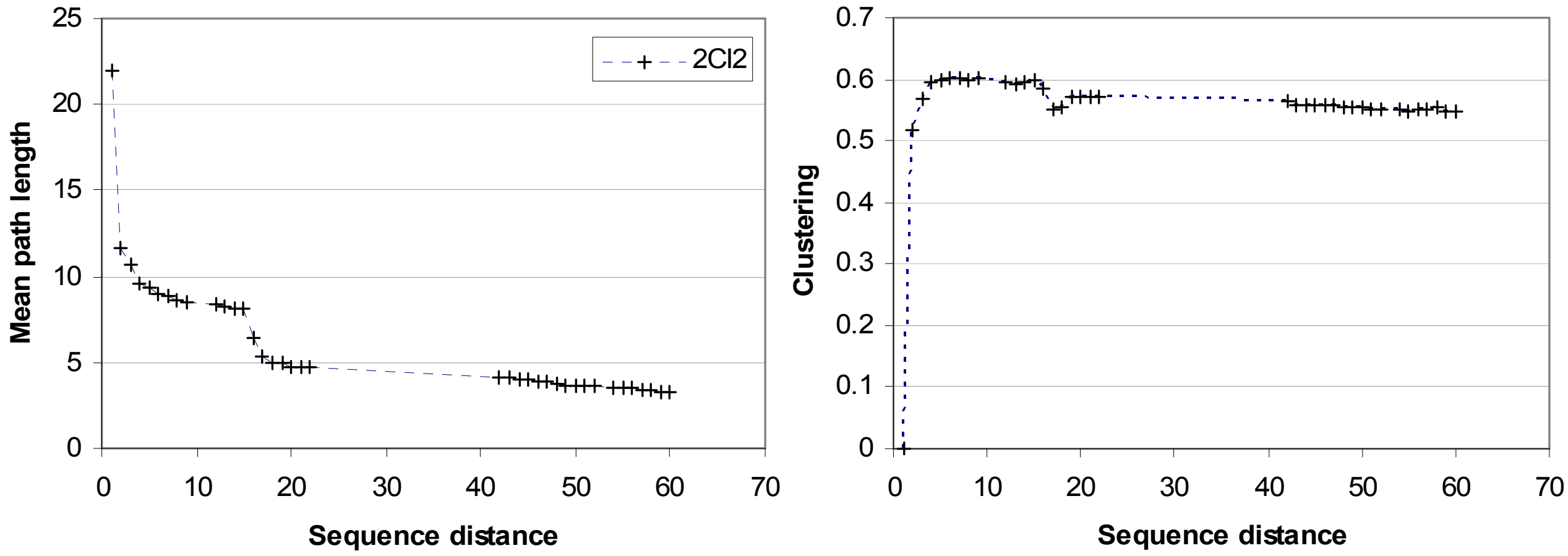

Fig. 1 Changes in mean path length and clustering as links are restored to the 2CI2 contact map or residue network in monotonically increasing sequence distance order. Mean path length and clustering are defined in the usual manner (section 2).

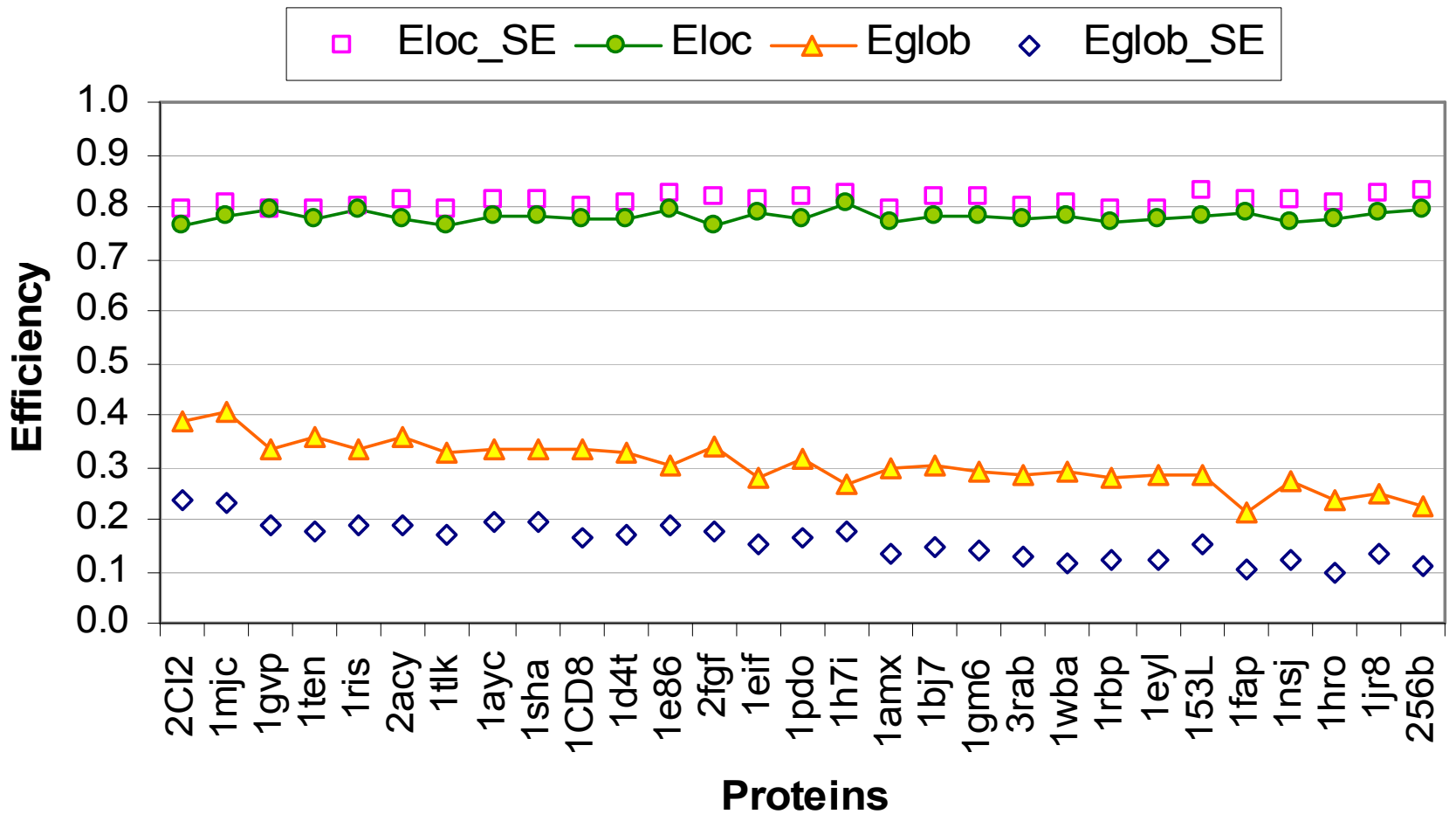

Fig. 2 Eglob is clearly above Eglob_SE showing that the addition of long-range links increases global efficiency. Eloc is only slightly below Eloc_SE showing that the addition of long-range links only slightly decreases local efficiency. The long-range links originate from the respective PRNs. Eglob_SE and Eloc_SE are measured on PCMs with only short-range links. Global and local efficiency are related to average path length and clustering respectively, and were proposed in [7] as an alternative way to characterize small-world networks. We use these measures because they are bounded within [0, 1] and can be conveniently placed in the same plot.

We would like to understand how the high levels of clustering are maintained given that the addition of a long-range edge, with everything else unchanged, decreases the clustering coefficients of the affected node pair and thus the average clustering coefficient of the network. The physicality of the protein molecule provides a partial answer: as atoms come close to each other so do their neighboring atoms. This detail was used in [6], but we desire a more principled approach.

## 2. Materials and Definitions

Protein contact maps (PCM) were constructed using the coordinates of the Cα atom of amino acids from the Protein Data Bank [8]. PCM ($i$, $j$) = 1 if the node pair ($i$, $j$) is situated less than 7Å from each other, otherwise PCM $(i, j) = 0$. A node pair $(i, j)$ is considered long-range if $|i - j| > 9$ [3]. Long-range links connect amino acids which are distant in the primary structure but are in close spatial proximity in the tertiary structure.

Define SPL($x, y$) as the length of a shortest path (in graph distance) between nodes $x$ and $y$. We assume that SPL($x, y$) = SPL($y, x$) to reduce computation time. The mean path length of a network G with N nodes is defined as APL(G) = $\frac{2}{N(N-1)} \sum_{i<j}^{N} SPL(i, j)$. SPL($x, y$) = 0 if no path exists between $x$ and $y$. The average graph distance between node $x$ and all other nodes in the network is defined as DIST($x$) = $\frac{1}{N-1} \sum_{i \neq x}^{N} SPL(x, i)$. The global efficiency of a network G is defined as Eglob(G) = $\frac{2}{N(N-1)} \sum_{i<j}^{N} \frac{1}{SPL(i, j)}$. $\frac{1}{SPL(x, y)} = 0$ if no path exists between $x$ and $y$ [7].

The clustering coefficient of a node $x$ is defined as CLUS($x$) = $\frac{2\, e_x}{k_x (k_x - 1)}$ where $k_x$ is the degree of node $x$, and $e_x$ is the number of links that exist amongst the $k_x$ nodes [4]. CLUS(G) = $\frac{1}{N} \sum_{i}^{N} CLUS(i)$. The local efficiency of a network G is defined as Eloc(G) = $\frac{1}{N} \sum_{i}^{N} Eglob(G_i)$ where $G_i$ is a subgraph of G comprising node $i$ and its direct neighbors [7]. The shortest graph distance between any pair of direct neighbors of a node is 1 (if the pair is directly connected) or 2.

**3. Method**

The Moore neighborhood of radius one surrounding cell $(i, j)$ in a PCM are the 9 closest cells to its North, South, East, West, NorthEast, SouthEast, SouthWest and NorthWest. We ignore cells which go beyond the boundary of a PCM. A neighborhood configuration comprises a binary sequence of length 9 which states how the set of 6 neighbor nodes $\{ i-1, i, i+1, j-1, j, j+1 \}$ are connected to each other. The inspected neighborhoods do not cross the main diagonal of a PCM.

The complete PCM for a protein is first built using the method in section 2. Then for each cell $(i, j)$ in the PCM where $j > (i + 2)$, we try to find from the $2^9$-1 possible combinations, a better or just as good neighborhood configuration different from the original one. We do this by substituting the original neighborhood configuration with an alternative configuration one at a time, while keeping all other parts of the PCM in the original form. The number of neighborhoods considered for a N × N PCM is (N-1)(N-2) / 2.

A neighborhood configuration $m$ is better or just as good as the original neighborhood configuration $m0$ if for every node $i$ in the neighborhood: DIST($i$) ≤ DIST0($i$) and CLUS($i$) ≥ CLUS0($i$) where $m0$ is used in the calculation of DIST0($x$) and CLUS0($x$), $m$ is used in the calculation of DIST($x$) and CLUS($x$), and $m \neq m0$. A neighborhood (configuration) is optimal if no better or just as good alternative configuration is found.

PRNs were randomized by rewiring their long-range links such that the number of links and the node degrees do not change. The effect of this randomization is to increase global efficiency and decrease local efficiency (Fig. 3).

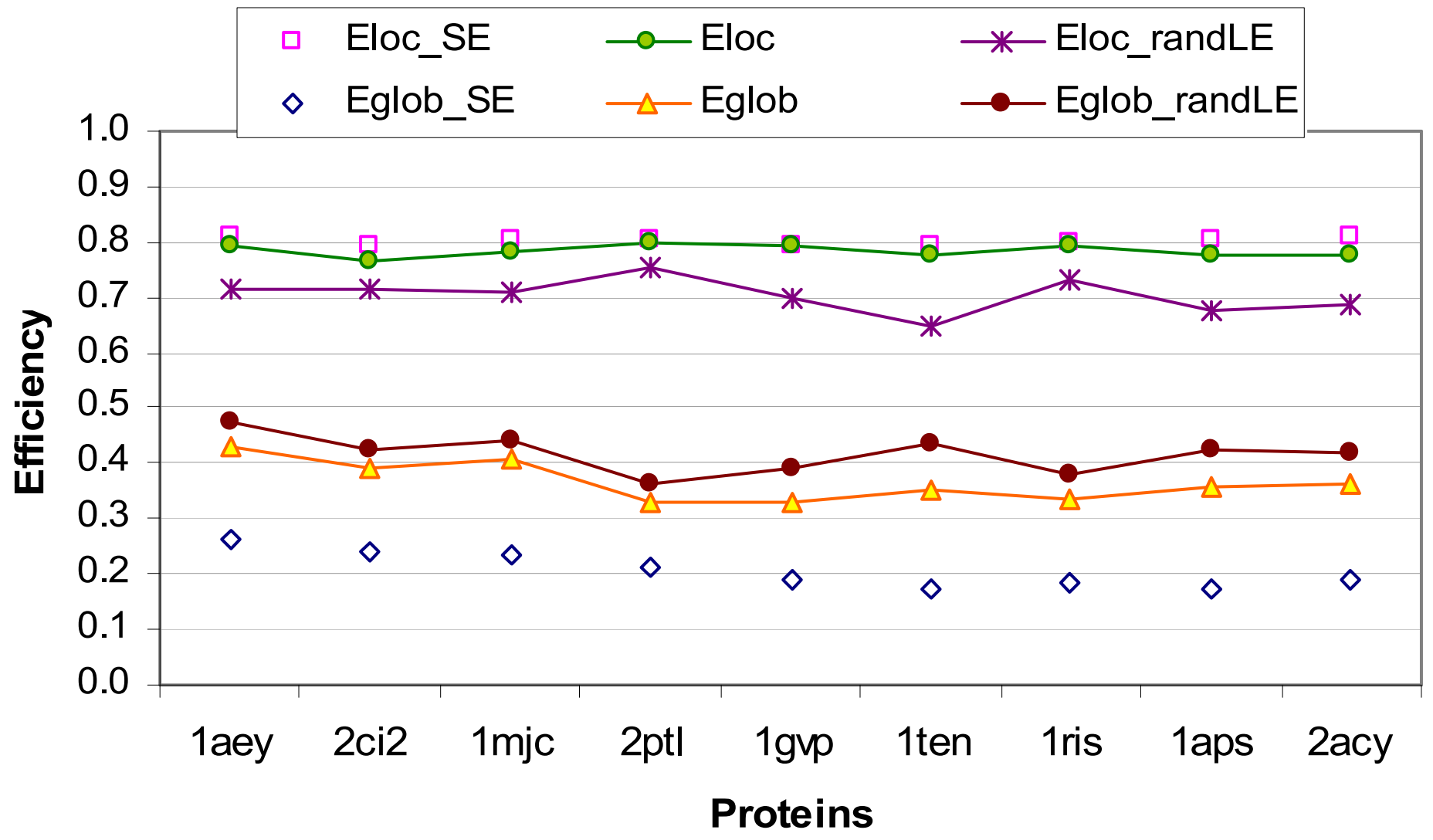

Fig. 3 Randomizing long-range links increases global efficiency (Eglob_randLE is above Eglob) and decreases local efficiency (Eloc_randLE is below Eloc). As in Fig. 2, Eloc is slightly below Eloc_SE and Eglob is clearly above Eglob_SE. Only chain A is used in 1aey, 2ptl and 1aps. This applies also to results in section 4.

## 4. Results

Fig 4. reports the percentage of neighborhoods considered in a protein's PCM which is not optimal. It shows that the Moore neighborhoods in PCMs are generally optimal in the sense described in section 3. That they are not 100% optimal may be a point for further study. Fig. 5 shows where the non-optimal neighborhoods are in four PCMs. In general, the non-optimal neighborhoods tend to clump together.

Fig. 6 reports the number of non-optimal neighborhoods for the non-randomized (left) and randomized (right) PCMs. Note the different scale on the y-axis. The x-axis in the bottom plot is labeled with the size of the corresponding protein considered in the top plot; so 1aey has a 58 × 58 PCM. By comparing the corresponding blue bars in the two plots, one finds that the number of non-optimal neighborhoods in randomized PCMs outnumber those in non-randomized PCMs. The red and yellow bars beside each blue bar represent the breakdown of the non-optimal neighborhoods into short-range and long-range respectively. The Moore neighborhood surrounding cell ($i$, $j$) is classified as short-range if $|i - j| \leq 9$, and long-range if $|i - j| > 9$.

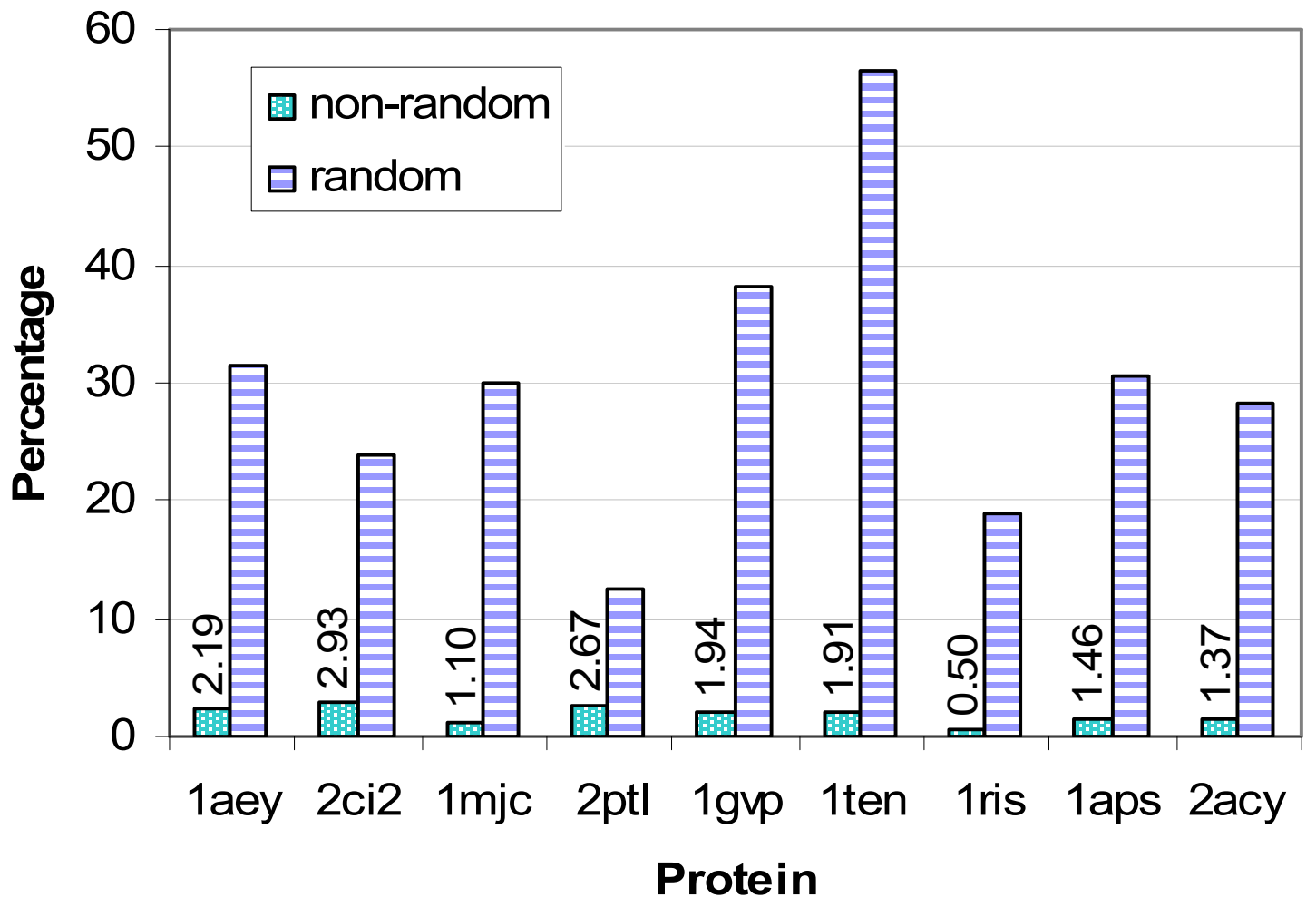

Fig. 4 Randomized PCMs have a larger percentage of neighborhoods which are non-optimal. Of the 2,016 neighborhoods inspected in the non-randomized 2CI2 PCM, 2.93% are non-optimal and the remaining 97.07% are optimal.

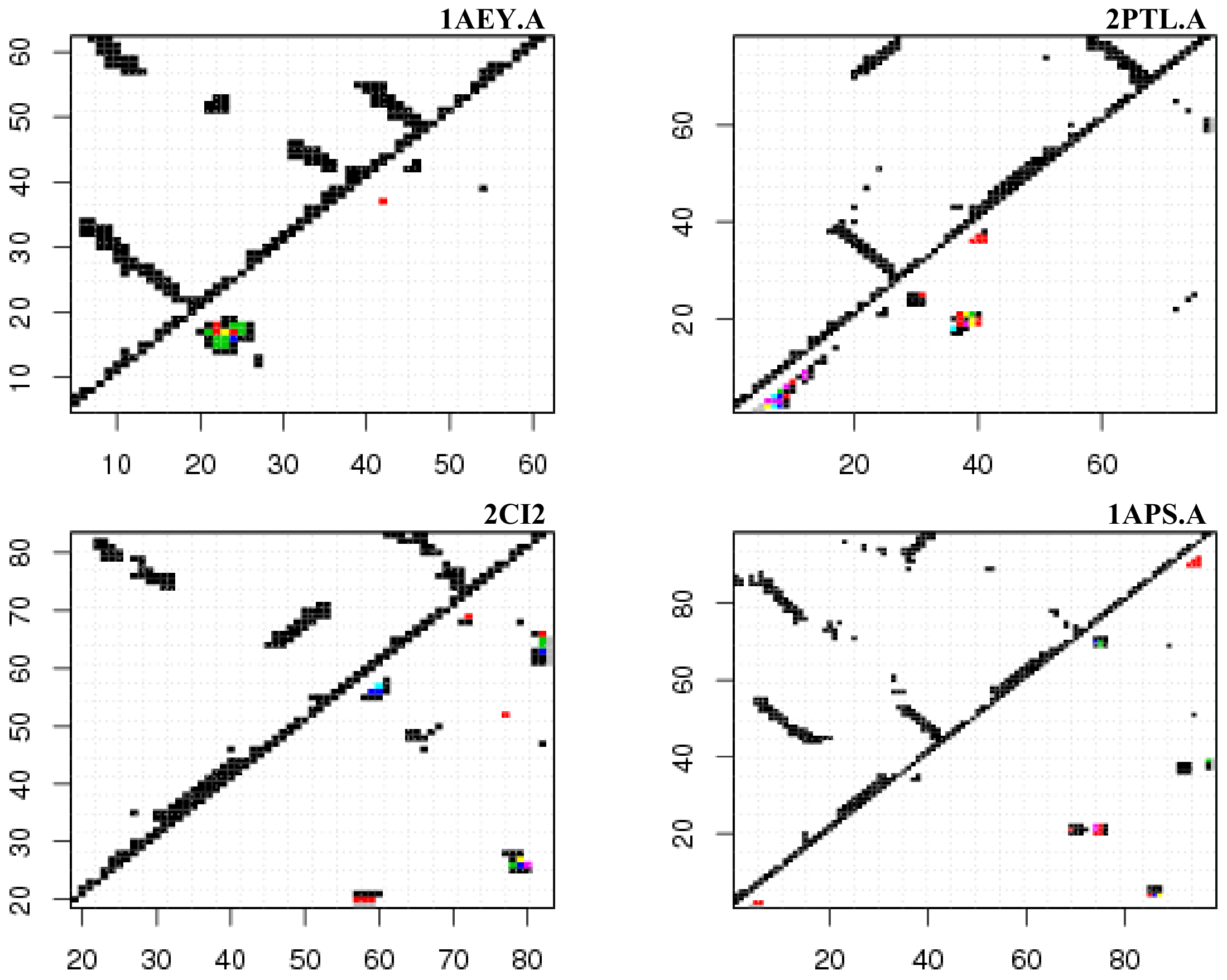

Fig. 5 A closer look at non-optimal neighborhoods. The PCM is shown in the upper-left triangle: the presence or absence of a link is indicated respectively by a black or white cell. The non-white cells in the lower-right triangle marks the cells with non-optimal neighborhood: different colors are used to emphasize that the number of alternative neighborhood configurations may vary from one or more. The x- and y- axes are labeled with the residue sequence number in the PDB file.

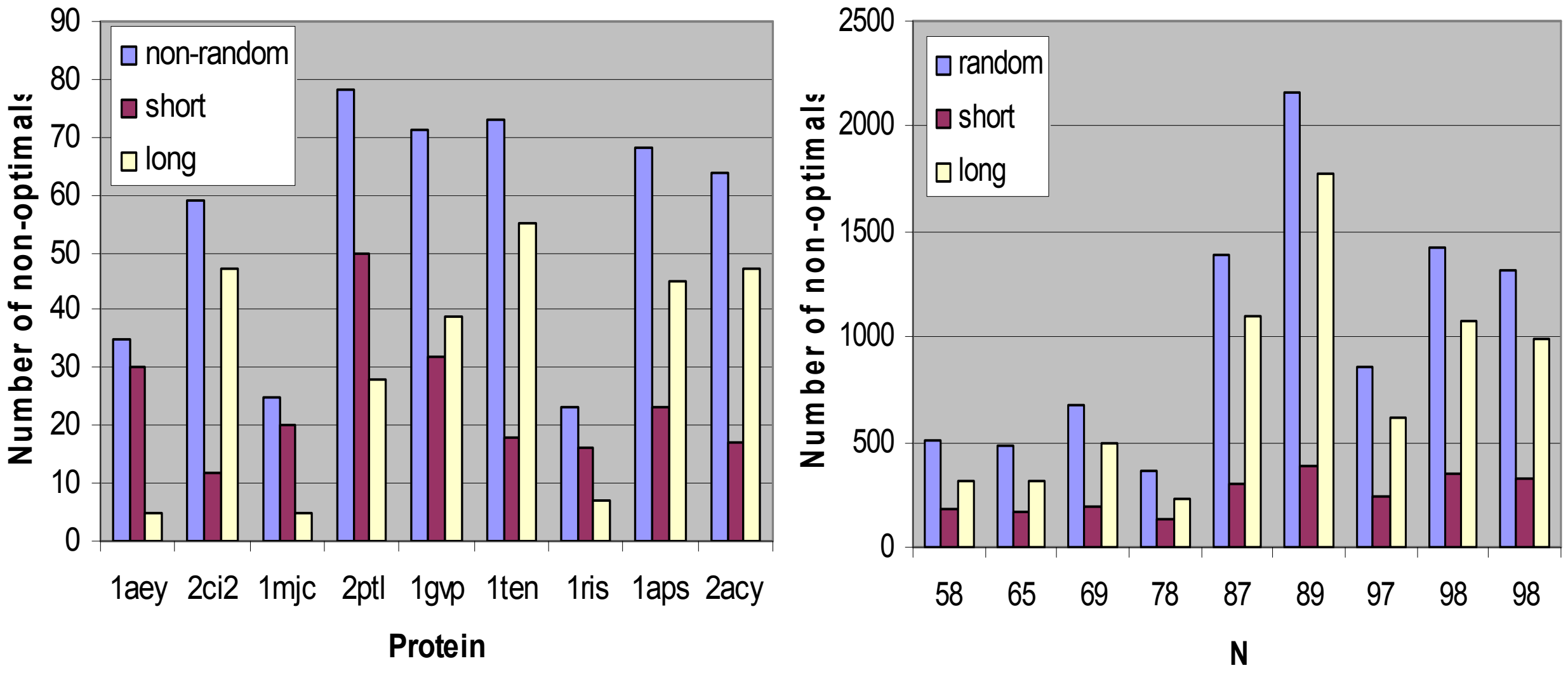

Fig. 6 Breakdown of non-optimal neighborhoods into short-range and long-range.

So far, we have examined optimality of Moore neighborhoods in complete PCMs, which are PRNs with their complete set of edges. In Fig.7 we report on optimality of neighborhoods as edges are added to four PCMs. The added edges come from the original PRNs and they are added in two ways: sorted according to increasing sequence distance, and unsorted. In both cases, the percentage of non-optimal neighborhoods generally decreases as more of a PCM's original edges are restored. However, a greater number of neighborhoods are optimal when the edges are restored in monotonically increasing sequence distance order than when the edges are restored in an unsorted manner.

## 5. Discussion

A question that arises from this result is how to use this information to "nudge" a non-optimal PCM towards optimality, or how to evolve optimal PCMs. An optimal PCM is one with little to no non-optimal neighborhoods. The problem of PCM prediction [9] can then be modified to one of maximizing the number of optimal neighborhoods. The original PCM or the one derived from existing PDB coordinates may be only one of a number of solutions. The fitness landscape of this maximization problem needs to be explored.

**Acknowledgements**

This work was made possible by the facilities of the Shared Hierarchical Academic Research Computing Network (SHARCNET:www.sharcnet.ca) and Compute/Calcul Canada.

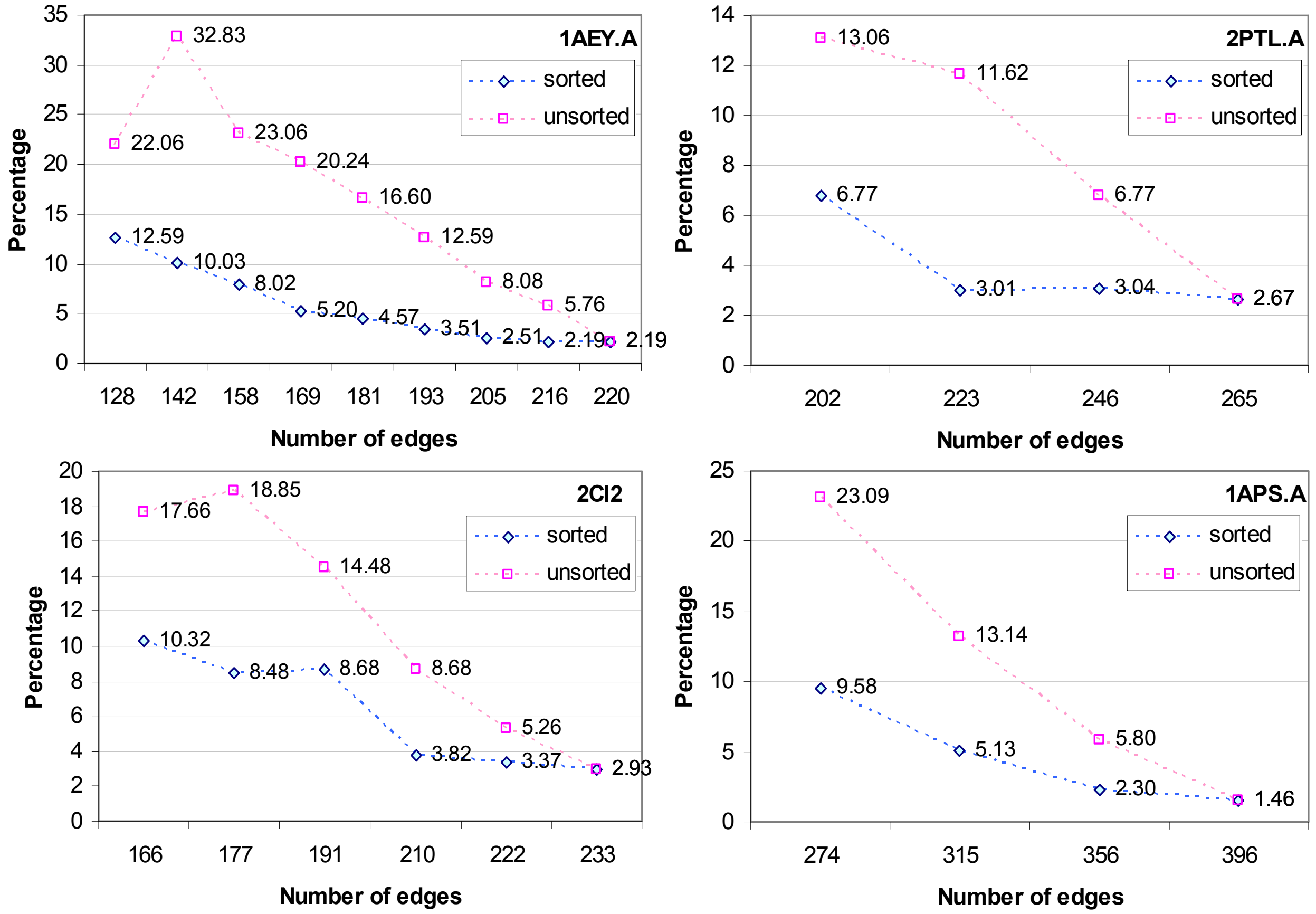

Fig. 7 Percentage of non-optimal neighborhoods as edges are restored to the 1AEY.A, 2CI2, 2PTl.A and 1APS.A PCMs. After 128 of the sorted edges are added, i.e. short edges are included before longer ones, 12.59% of neighborhoods in the 1AEY.A PCM are non-optimal.